\documentclass[preprint,prd,noshowpacs,nofootinbib]{revtex4}
\usepackage{amssymb}
\usepackage{amsmath}
\usepackage{graphicx}

\begin{document}

\title{Aharonov-Bohm phase for an electromagnetic wave background}

\author{Max Bright}
\email{neomaxprime@mail.fresnostate.edu}
\affiliation{Department of Physics, California State University Fresno, Fresno, CA 93740-8031, USA}

\author{Douglas Singleton}
\email{dougs@csufresno.edu}
\affiliation{Department of Physics, California State University Fresno, Fresno, CA 93740-8031, USA 
and 
ICTP South American Institute for Fundamental Research,
UNESP - Univ. Estadual Paulista, S{\~a}o Paulo, SP, Brasil 01140-070}

\author{Atsushi Yoshida}
\email{ay9a@virginia.edu}
\affiliation{Department of Physics, University of Virginia, Charlottesville, VA 22904-4714, USA 
and Hue University College of Education, Hue, Vietnam}
\date{\today}

\begin{abstract}
The canonical Aharonov-Bohm effect is usually studied with time-independent potentials. In this work, we 
investigate the Aharonov-Bohm phase acquired by a charged particle moving in {\it time-dependent} potentials . In particular, 
we focus on the case of a charged particle moving in the time varying field of a plane electromagnetic wave. We 
work out the  Aharonov-Bohm phase using both the potential ({\it i.e.} $\oint A_\mu dx ^\mu$) and field ({\it i.e.} 
$\frac{1}{2}\int F_{\mu \nu} d \sigma ^{\mu \nu}$) forms of the Aharanov-Bohm phase. We give conditions in terms of the 
parameters of the system (frequency of the electromagnetic wave, the size of the space-time loop, amplitude of the 
electromagnetic wave) under which the time varying Aharonov-Bohm effect could be observed. 
\end{abstract}

\maketitle
\section{Introduction}
In this work, we investigate the Aharonov-Bohm phase difference picked up by charged particles that go around a closed space-time 
loop in the presence of the time-dependent potentials and fields of an electromagnetic plane wave. 

Theoretical investigations of the Aharonov-Bohm effect \cite{AB,ES} generally involve time-independent potentials. The canonical 
example of the Aharonov-Bohm effect is that of charged particles in a two-slit experiment with an infinite solenoid,
carrying a constant magnetic flux, placed between the slits. Each charged particle picks up 
an additional phase due to the non-zero vector potential outside the solenoid, even though the electric 
and magnetic fields outside the solenoid are zero. The experimental tests of the Aharonov-Bohm effect 
have also generally been done with time-independent fields \cite{chambers, tonomura}. 

In contrast to the time-independent Aharonov-Bohm effect, there have been only a few theoretical studies 
of the {\it time-dependent} Aharonov-Bohm effect. Some of the works \cite{kampen, roy, brown, gaveau, singleton, singleton2} 
have focused on a solenoid with a time-dependent magnetic flux, while others \cite{chiao, ford} have used 
an electromagnetic plane wave to obtain time-dependent potentials and fields. 

On the experimental side, there are only two cases that we have 
found where the time-dependent Aharonov-Bohm effect was tested experimentally.
The first test was an accidental experiment by Marton {\it et al.} \cite{marton} where an electron two-slit interference 
experiment was set up and the interference pattern was observed. However, it was later determined that the region through 
which the electrons traveled was contaminated with a 60 Hz magnetic field of unknown strength. The question in regard to the 
results of the accidental experiment in \cite{marton} are: ``Why was the interference pattern seen at all? Why did it not shift 
back a forth at 60 Hz?" At first, it was thought that the result of Marton {\it et al.} 
was evidence against the Aharonov-Bohm effect. However, two explanations have 
been put forward as to why the experiment in \cite{marton} saw a static interference pattern. In \cite{brill},
the idea was advanced that the time varying Aharonov-Bohm phase was compensated for by a phase shift coming from the
direct ${\bf v \times B}$ force which acted on the electrons. In \cite{singleton3}, the explanation given for the observation of 
the static interference pattern in \cite{marton} was due to a cancellation between the time varying Aharonov-Bohm phase and 
a phase coming from the induced electric field that accompanied the time varying magnetic field. The second test of the time-dependent
Aharonov-Bohm was the experiment in \cite{chentsov, ageev}. This experiment used fields from an electromagnetic wave 
with a frequency in the microwave region and was along the lines of the set-up suggested in \cite{chiao} for testing the
time varying Aharonov-Bohm effect. This experiment was also along the lines of \cite{ford}, which studied decoherence effects
due to the time varying Aharonov-Bohm phase coming from an electromagnetic wave. The results of the experiment 
described in \cite{chentsov, ageev} were that evidence for an Aharonov-Bohm phase from the time varying fields and potentials 
was not observed -- thus these results were similar
to the accidental experiment of Marton {\it et al.}, where the effect of the time variation was not seen in the interference 
pattern. The explanation of the non-observation of the time variation in the experiment \cite{chentsov} was that the parameters of
the set-up were such that the time variation effect was too small to be seen \cite{ageev}. We come to a similar conclusion from our 
analysis -- in order to observe the time variation one must carefully chose the various parameters of the set-up: 
the frequency and amplitude of the electromagnetic wave, the size of the loop, the velocity of the particle, 
{\it etc.} In the conclusion, we give conditions under which one might see evidence of the time varying Aharonov-Bohm effect.

There are two points to make before we move on to our detailed analysis. First, the time-dependent Aharonov-Bohm effect is invariably a
type II Aharonov-Bohm effect. The type I Aharonov-Bohm effect is when the charged particle develops a phase
while moving through a region that is free of electric and magnetic fields, as in the original time-independent, infinite solenoid set-up. 
The type II Aharonov-Bohm effect is when  the charged particle develops an Aharonov-Bohm phase, but while 
moving through a region of space where the fields are not zero. A
typical example of a type II effect is the Aharonov-Casher effect \cite{AC}, where a neutral particle with a magnetic moment moves through an
electric field and, in doing so, picks up a Aharonov-Bohm-like phase. Second, the set-up we study here is a combination and generalization of
the time-dependent Aharonov-Bohm effect studied in \cite{chiao} and \cite{ford}. In particular for our set-up, both the electric and magnetic 
Aharonov-Bohm effects are non-zero. In reference \cite{chiao}, the set-up was taken so that only the time varying magnetic field gave an
Aharonov-Bohm phase, while in reference \cite{ford}, the set-up was such that only the time varying electric field gave a non-zero 
contribution to the Aharonov-Bohm phase. Further, these two prior works calculated the Aharonov-Bohm phase in different ways: 
reference \cite{chiao} used the line integral of the potentials to obtain the phase while reference \cite{ford} used the 
area integral of the fields. Here, we calculate the phase using both the line integral of the potentials and the area integral of the fields. This provides an explicit working out of Stokes' theorem in 4D. In the literature, we have found no examples of this, although there are of course plenty of explicit worked out examples of Stokes' theorem in 3D.

\section{General set-up of potentials, fields and path}
\label{potentials}

In this section, we give the set-up for the potentials, fields and path that we will use. We consider a linearly polarized
plane wave traveling along the $z$-axis in either the $+$ or $-$ direction. The covariant, four-vector potential for this is
\begin{equation}
\label{Au}
A_\mu = \left( 0 , A_0 f(\omega t \pm k z) , 0, 0 \right) ~, 
\end{equation}
where $A_0$ is the amplitude and $\omega , k$ are the frequency and wave number for the wave. The $(+)$ sign is a
wave traveling in the $-z$ direction and the $(-)$ is a wave traveling in the $+z$ direction. The electric and magnetic fields
can be obtained from \eqref{Au} using $F_{\mu \nu} = \partial _\mu A_\nu - \partial _\nu A_\mu$ with the result
\begin{equation}
\label{EB-wave}
F_{01} = E_x = A_0 \frac{\omega}{c} f' (\omega t \pm k z) ~~~;~~~ F_{13} = B_y = \mp A_0 k f' (\omega t \pm k z) ~,
\end{equation}
where prime means a derivative with respect to the argument of $f$, namely
$\zeta _\pm = \omega t \pm k z$. The electromagnetic wave in 
\eqref{EB-wave} is polarized in the $x$ direction. We could consider a more complicated wave traveling in the $z$-direction, with 
the electric and magnetic fields having components in both $x$ and $y$ directions. The vector potential for such wave would have the
form $A_\mu = \left( 0 , A_0 f(\omega t \pm k z + \varphi_1) , B_0 f(\omega t \pm k z +\varphi_2) , 0 \right)$, where $A_0, B_0$ 
are amplitudes and $\varphi _1 , \varphi _2$ are phases. For our purposes, the four-potential in \eqref{Au} is sufficient since for the closed loop path that we pick, with no motion of the particles in the $y$ direction, only the $x$ component of the 
four-vector potential will contribute. If we had picked our loop to be in the $yz$ plane, then it would be a vector 
potential of the form  $A_\mu = \left( 0 , 0 , B_0 f(\omega t \pm k z) , 0 \right)$ which would give a non-zero contribution. 

\begin{figure}
  \centering
	\includegraphics[trim = 0mm 0mm 0mm 0mm, clip, width=6.0cm]{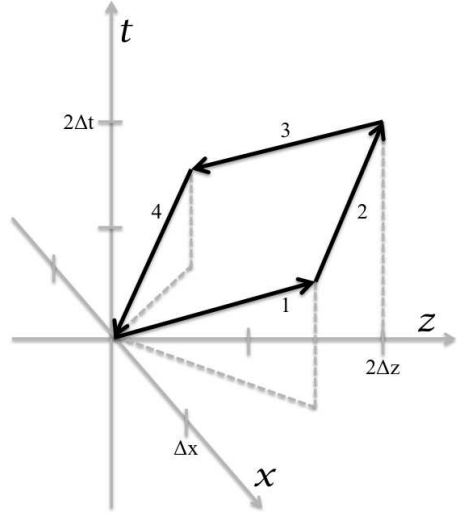}
\caption{{\it The full space-time loop in $txz$.}}
\label{fig1}
\end{figure}
\begin{figure}
  \centering
	\includegraphics[trim = 0mm 0mm 0mm 0mm, clip, width=6.0cm]{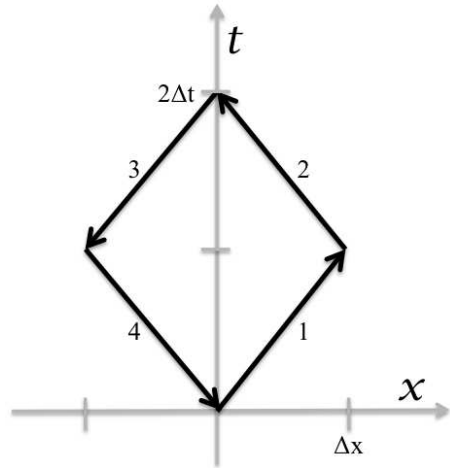}
\caption{{\it The projection of the space-time loop in $tx$ plane.  This also represents the ``upright diamond" which we will discuss in the next section.}}
\label{fig2}
\end{figure}
\begin{figure}
  \centering
	\includegraphics[trim = 0mm 0mm 0mm 0mm, clip, width=6.0cm]{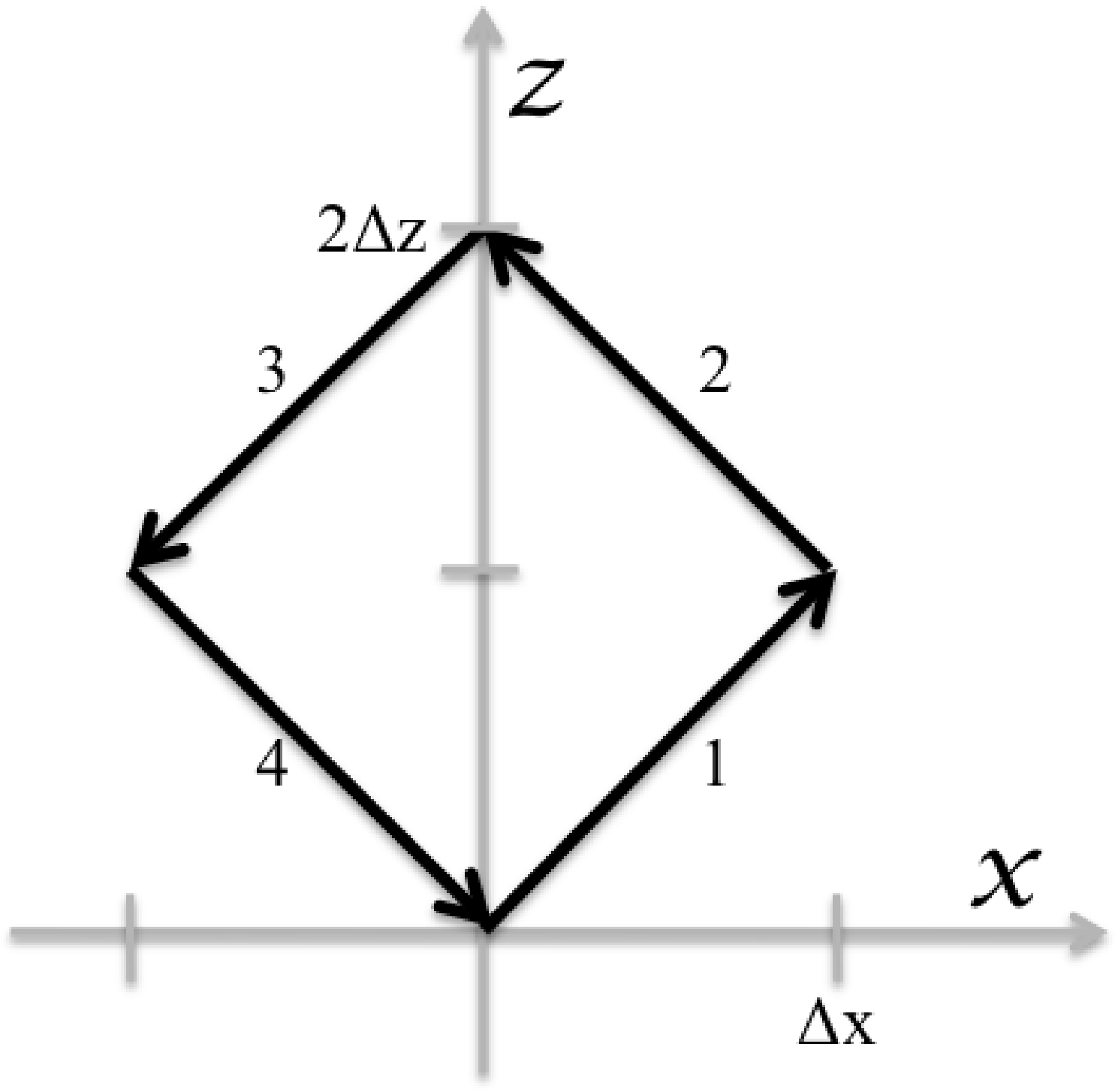}
\caption{{\it The projection of the space-time loop in $xz$ plane.}}
\label{fig3}
\end{figure}

The closed space-time path that we chose in evaluating the loop integral $\oint A_\mu dx^\mu$ is shown fully in figure \ref{fig1}. 
The projections in the $tx$ and $zx$ planes are shown, respectively, in figure \ref{fig2} and figure \ref{fig3}. In the experiment, the charged particles would travel both paths $1$ and $2$, and $3$ and $4$, in the forward 
time-direction, {\it i.e.} the charged particles would leave the origin and travel forward along $1$ and $2$ and also forward 
along $3$ and $4$, with each picking up some different Aharonov-Bohm phase along these paths. To take the phase 
{\it difference} between path $1$ and $2$ versus path $3$ and $4$, one puts a negative sign in front of the integral along path
$3$ and $4$ (or as well one could put the negative sign in front of the integrals along path $1$ and $2$, thus reversing the 
direction of the space time loop in figure \ref{fig1}). In this way, one ends with a closed space-time loop integral. 
This is what is done in the usual static Aharonov-Bohm analysis, but with purely spatial loop integrals.  

The paths start from the origin, $x=z=t=0$, and then travel with velocity ${\bf v} = (\pm v_x, 0, v_z)$, for a time $\Delta t$, to
the two spatial points $(\pm \Delta x, 0, \Delta z)$. This gives paths $1$ and $4$.  From these two points, one traces 
back to the point $(0, 0, 2 \Delta z)$ in a time $\Delta t$ (this means a total time of $2 \Delta t$); 
this gives paths $2$ and $3$. Note, the total time from the origin to $(0, 0, 2 \Delta z)$  is $2 \Delta t$, 
so $\Delta t = \frac {\Delta z}{v_z} = \frac{\Delta x}{v_x}$. Now, for the explicit evaluation of 
$\oint A_\mu dx^\mu$ in the next section, we need to give the equations describing each path:
\begin{eqnarray}
\label{path1}
{\rm Path ~ 1} ~~ (0 < t < \Delta t): ~ t&=& \frac{x}{v_x} = \frac{z}{v_z} ~, \\
\label{path2}
{\rm Path ~ 2} ~~ (\Delta t < t < 2\Delta t): ~ t&=& -\frac{x}{v_x} + 2\Delta t ~~ = ~ \frac{z}{v_z} ~,\\
\label{path3}
{\rm Path ~ 3} ~~ (\Delta t < t < 2\Delta t): ~ t&=& \frac{x}{v_x} + 2\Delta t ~~ = ~ \frac{z}{v_z} ~,\\
\label{path4}
{\rm Path ~ 4} ~~ (0 < t < \Delta t): ~ t&=& -\frac{x}{v_x} ~~ = ~ \frac{z}{v_z} ~.
\end{eqnarray}

The magnitude of the velocity along any path is $v = \sqrt{ v_x ^2 + v_z^2} \le c$.   For the projections of the full space-time path from figure \ref{fig1} into the 
$tx$ plane in figure \ref{fig2}, the slopes of the paths are greater than 1 since  $v_x, v_z<c$. 

For the Aharonov-Bohm effect with a plane wave background, both the electric and magnetic Aharonov-Bohm effects can be non-trivial at the same time.  For the usual static magnetic Aharonov-Bohm effect, the line integral of the the 3-vector potential, 
${\bf A}$, is related to the surface area of the magnetic field via Stokes' theorem in 3D, 
$\oint {\bf A \cdot} d {\bf x} = \int {\bf B \cdot} d{\bf a}$. In order for this magnetic Aharonov-Bohm phase to be non-zero,
the magnetic field, ${\bf B}$, must have a component along the area normal direction, $d {\bf a}$. For 
our space-time area, the spatial projection is in the $xz$ plane, which therefore has a spatial area direction in the 
$y$-direction. This then gives a non-zero contribution for the magnetic field of the plane wave since ${\bf B} \propto {\hat {\bf y}}$. 
For the usual static electric Aharonov-Bohm effect, it is the time integral of the scalar potential, $\phi$, which is 
important (${\it i.e.} \int \phi dt$). Using ${\bf E} = - \nabla \phi$, one finds that the electric Aharonov-Bohm phase can be written
as $\int \phi dt = -\int {\bf E \cdot} d{\bf x} dt$. Note that the spatial area of the magnetic case, $d{\bf a}$, is replaced by a
space-time area $d{\bf x} dt$. In order to get an electric contribution to the Aharonov-Bohm phase, the electric field must have
a component along the direction of the path. 

For the wave traveling in the $z$-direction, and for the area of the loop which has a projection in the $xz$ plane, both the magnetic field
in the $y$-direction and the electric field in the $x$-direction will give non-zero contributions to the time varying Aharonov-Bohm phase. 
In the case studied in \cite{ford}, the wave was taken to be traveling in the $y$-direction, the electric field was polarized in the
$z$-direction, and the magnetic field was in the $x$-direction. The loop chosen in \cite{ford} 
was also in the $xz$ plane, so, in that situation, there was only a non-zero electric Aharonov-Bohm phase; 
the magnetic Aharonov-Bohm phase was zero. In our set-up {\it both} the magnetic and electric Ahronov-Bohm phases play a role.

\section{Aharonov-Bohm phase via the potentials}
\label{AB-potentials}

Using the above set-up, we now calculate $\oint A_\mu dx^\mu$, which gives the Aharonov-Bohm phase when multiplied by 
$\frac{e}{\hbar c}$. We will calculate the phase picked up along the four path lengths
given in figure \ref{fig1} and then add them. The electromagnetic wave has a frequency and wave 
number given by $\omega$ and $k$, which satisfy $\frac{\omega}{k} =c$.
The velocity on any given leg of the path in figure \ref{fig1} is ${\bf v} = (\pm v_x, 0, v_z)$.

In carrying out the calculation for the loop integral for the case in figure \ref{fig1}, we will first calculate $\oint A_\mu dx^\mu$
in the frame where $v_z=0$; this is the ``upright diamond" loop of figure \ref{fig2}. Then, to obtain the more general case, we will
boost the result of the upright diamond so that the particle develops a velocity, $v_z$, in the $+z$ direction.  \\

{\bf The ``upright diamond"}: For the upright diamond in figure \ref{fig2}, we have $z = 0$ and ${\bf v} = (\pm v_0, 0, 0)$, and 
we take the wave number and frequency as $k_0$ and $\omega_0$, respectively. The four paths are parametrized as:
\begin{eqnarray}
\label{path1ud}
{\rm Path ~1} ~~ (0 < t < \Delta t): ~ t&=& \frac{x}{v_0} ~,\\
\label{path2ud}
{\rm Path ~2} ~~ (\Delta t < t < 2\Delta t): ~ t&=& -\frac{x}{v_0} + 2\Delta t ~,\\
\label{path3ud}
{\rm Path ~3} ~~ (\Delta t < t < 2\Delta t): ~ t&=& \frac{x}{v_0} + 2\Delta t ~,\\
\label{path4ud}
{\rm Path ~4} ~~ (0 < t < \Delta t): ~ t&=& -\frac{x}{v_0} ~.
\end{eqnarray}

Along each path, since $z=0$, $f(\omega_0 t \pm k_0z)$ simplifies to $f(\omega_0 t)$.  For path~1, from \eqref{path1ud}, 
we find
\begin{equation}
\label{path1a}
\int _1 A_\mu dx^\mu =  A_0 \int _0 ^{\Delta x} f\left(\frac{k_0}{\beta_0}x \right) dx   =  \frac{A_0 \beta_0}{k_0} 
\left[ F \left( \Delta \zeta_0 \right) - F(0) \right] =  A_0 \Delta x ~\langle f(\zeta) \rangle_{\rm I}~,
\end{equation}
where $\beta_0 = v_0/c$, $c k_0 = \omega_0$, and $F(\zeta) = \int f(\zeta ) d \zeta$ is the integral function of $f$. 
$\Delta \zeta_0$ is the phase shift up to the half-way point,
\begin{eqnarray}
\Delta \zeta_0 = \frac{k_0}{\beta_0} \Delta x = \omega_0 \Delta t~.
\end{eqnarray}
In the final expression in \eqref{path1a} we have written the result in terms of
$\langle f(\zeta) \rangle_{\rm I} \equiv \frac{1}{\Delta \zeta_0} \int _0 ^{\Delta \zeta_0} f(\zeta ) d \zeta$, 
which is the average of $f(\zeta)$ in the interval ${\rm I} = \left\{ \zeta : 0 < \zeta < \Delta \zeta_0 \right\}$. 

For path $2$, we have, from \eqref{path2ud},  
\begin{eqnarray}
\label{path2a}
\int _2 A_\mu dx^\mu &=&  A_0 \int ^0 _{\Delta x} f \left( -\frac{k_0}{\beta_0} x  + 2\omega_0 \Delta t \right) dx \nonumber \\
&=&  - \frac{A_0 \beta_0}{k_0}\left [F\left( 2 \omega_0 \Delta t \right) - F\left(-\frac{k_0}{\beta_0} \Delta x + 2\omega_0 \Delta t \right) \right] ~  \\
&=&  - \frac{A_0 \beta_0}{k_0}\left [ F\left( 2\Delta \zeta_0 \right) - F\left(\Delta \zeta_0 \right) \right] 
=  - A_0 \Delta x ~\langle f(\zeta) \rangle_{\rm II} \nonumber ~,
\end{eqnarray}
where $\langle f(\zeta) \rangle_{\rm II}$ is the average of $f(\zeta)$ over the interval
${\rm II} = \left\{ \zeta : \Delta \zeta_0 < \zeta < 2\Delta \zeta_0 \right\}$.

In a similar manner, for path $3$ and path $4$, one finds 
\begin{eqnarray}
\int _3 A_\mu dx^\mu = \int _2 A_\mu dx^\mu \ ; \ \ \ \int _4 A_\mu dx^\mu = \int _1 A_\mu dx^\mu ~.
\end{eqnarray}
Altogether, the upright diamond loop integral is 
\begin{eqnarray}
\label{upright1}
\oint A_\mu dx^\mu &=& \int _1 A_\mu dx^\mu + \int _2 A_\mu dx^\mu + \int _3 A_\mu dx^\mu + \int _4 A_\mu dx^\mu \nonumber \\
&=& 2\frac{A_0 \beta_0}{k_0}  \left(  \left[ F \left( \Delta \zeta_0 \right) - F(0) \right] - \left [ F\left( 2\Delta \zeta_0 \right)
- F\left( \Delta \zeta_0 \right) \right] \right)  \\
&=& 2\frac{A_0 \beta_0}{k_0} \left( 2F\left(\Delta \zeta_0 \right) -  F(0) - F\left( 2\Delta \zeta_0 \right)  \right)
\nonumber ~,
\end{eqnarray}
or in terms of the avergaes of $f(\zeta)$,
\begin{eqnarray}
\label{upright2}
\oint A_\mu dx^\mu = 2A_0 \Delta x \left( \langle f(\zeta) \rangle_{\rm I}  -  \langle f(\zeta) \rangle_{\rm II} \ \right).
\end{eqnarray}
The loop integral is zero if the average of the vector potential in the first half (I) 
and the second half (II) are the same.\\

{\bf Boost along $z$-direction}: In order to evaluate the loop integral for $v_z \ne 0$, we now boost the frame along the
$z$-direction by $\beta$. This gives the particle a velocity in the $z$ direction, and changes the 
upright diamond of figure \ref{fig2} to the tilted diamond of figure \ref{fig1}. In equation 
\eqref{upright1}, the only parameters that change are the wave number $k_0$ and the $x$-component of velocity, $\beta_0$.  
$\Delta \zeta_0$ also transforms due to $k_0$ and $\beta_0$ changing.  Te parameters $\Delta x$ and $A_0$ do not change
since they are perpendicular to the boost direction.  After the boost,
\begin{equation}
\label{k}
\beta_0 \ \rightarrow \ \beta_x = \sqrt{1 - \beta^2} \beta_0 = \beta_0/\gamma ~~~;~~~
k_0 \ \rightarrow \ k = \sqrt{\frac{1 \pm \beta}{1 \mp \beta}} \ k_0 ~.
\end{equation}
In equation \eqref{upright1}, only the combination $\frac{k_0}{\beta_0}$ shows up. Using \eqref{k} we find
\begin{equation}
\frac{k_0}{\beta_0} \rightarrow \sqrt{\frac{1 \mp \beta}{1 \pm \beta}} \ k  \cdot \frac{1}{\gamma \beta_x} = (1 \mp \beta) \frac{k}{\beta_x}
\end{equation}

Setting $\beta = -\beta_z = -v_z/c$, we get the final result 
\begin{eqnarray}
\label{pathtot}
\oint A_\mu dx^\mu &=& \frac{2A_0}{k} \frac{\beta_x}{1 \pm \beta_z} \left[ 2F\left(\frac{1 \pm \beta_z}{\beta_x} k\Delta x \right)
- F(0) - F\left( 2 \frac{1 \pm \beta_z}{\beta_x} k\Delta x \right)  \right] \nonumber \\
&=& \frac{2A_0}{k'} \left[ 2F\left(k'\Delta x \right) -  F(0) - F\left( 2k'\Delta x \right) \right]  \\
&=& \frac{2A_0}{k'}  \left[  2F(\Delta \zeta) -  F(0) - F(2\Delta \zeta)  \right] \nonumber ~,
\end{eqnarray}
where $k'$ is defined as
\begin{eqnarray}
\label{kprime}
k' = \frac{1 \pm \beta_z}{\beta_x} k ~,
\end{eqnarray}
and $\Delta \zeta$ is the phase shift to the half-way point
\begin{eqnarray}
\label{zeta}
\Delta \zeta = k'\Delta x = \omega (1\pm \beta_z) \Delta t = \omega \Delta t \pm k \Delta z ~.
\end{eqnarray}

The last line of equation \eqref{pathtot} is the main result of this 
section and we want to make some comments/remarks about its physical meaning. First, we can look at an expansion 
of the function $F(\eta) = F(0) + \eta F'(0) + \frac{1}{2} \eta ^2 F''(0) +...$, where $\eta = \Delta \zeta = k' \Delta x$ or
$\eta = 2 \Delta \zeta = 2 k' \Delta x$, depending on which term in \eqref{pathtot} one is expanding. 
Using this, one finds that \eqref{pathtot} becomes
\begin{equation}
\label{pathtot1}
\oint A_\mu dx^\mu = - 2 A_0 k ' \Delta x ^2 F '' (0) + {\cal O}( (\Delta \zeta)^3) ~.
\end{equation}
Thus, to first order in $\Delta \zeta$, the time varying Aharonov-Bohm phase vanishes. This result can be compared to the result in 
\cite{singleton} \cite{singleton2}, where the authors found that for a solenoid with a time varying flux, the time-dependent 
part of the Aharonov-Bohm phase vanishes to first order as well. 

Next, if we take the particle speed to be much less than the speed of light, 
$v = \sqrt{v_x^2 + v_z^2} \ll c$, then, from \eqref{kprime}, $k ' \approx k/\beta_x$.  Hence the condition 
$k'\Delta x \ll 1$ is identical to ${\Delta x}/\beta_x \ll \lambda$, or equivalently $\Delta t \ll T = \frac{2 \pi}{\omega}$, 
where $\lambda$ and $T$ are the wavelength and the period of the electromagnetic wave.  Since the approximate vanishing of 
$\oint A_\mu dx^\mu$ requires $k ' \Delta x \ll 1$, the condition above imposed on the wavelength or the period 
of the electromagnetic wave is a condition for the approximate vanishing of the loop integral.  
At the end of the next section, we will comment more on the use of the result in \eqref{pathtot1} to determine 
when the time-dependent Aharonov-Bohm effect is important/observable.

One final question we want to address: ``Under what conditions does $\oint A_\mu dx^\mu$ 
vanish exactly".  According to \eqref{pathtot} and \eqref{pathtot1}, if $k' \rightarrow 0$, the loop 
integral vanishes, which in turn means $\beta_z \rightarrow \mp 1$ according to \eqref{kprime}.  
Thus, if the particle traverses the loop at close to $c$ and moves in the same direction as the electromagnetic wave 
(in this case the $z$ -direction) the time-dependent Aharonov-Bohm phase will vanish. This was the same conclusion 
reached for the time dependent, non-Abelian Aharonov-Bohm effect \cite{bright} using the time-dependent, non-Abelian 
plane waves of Coleman \cite{coleman}.  However, in the present case, due to ‎the constraint 
$\sqrt{\beta_x^2 + \beta_z^2} \le 1$, $\beta_z \rightarrow 1$ means $\beta_x \rightarrow 0$, 
the $x$-component of the velocity is vanishingly small, and that the diamond shaped loop becomes elongated in the $z$
direction since the ratio of the lengths goes as $\Delta x/\Delta z \rightarrow 0$.

There is another way that the loop integral vanishes exactly.  Rewriting \eqref{pathtot} in terms of the original 
function $f(\omega t \pm kz)$, we have
\begin{eqnarray}
\label{ave-phase}
\oint A_\mu dx^\mu = 2A_0 \Delta x \left( \langle f(\zeta) \rangle_{\rm I}  -  \langle f(\zeta) \rangle_{\rm II} \right) ~,
\end{eqnarray}
where the notation is similar to equation \eqref{upright2} and with the intervals defined as 
\begin{eqnarray}
{\rm I} = \{ \zeta : \ 0 < \zeta < \Delta \zeta \ \} ; \ \ \ {\rm II} = \{ \zeta : \Delta \zeta < \zeta < 2\Delta \zeta \ \} ~,
\end{eqnarray}
where $\Delta \zeta = \omega \Delta t \pm k \Delta z$, the phase change to the half-way point.
This means that if the average of the vector potential experienced by the particle in the first half, 
$0 < t < \Delta t$, is the same as the average of that experienced in the latter half, 
$\Delta t < t < 2 \Delta t$, then the loop integral vanishes.  This happens, for example, if $f(\zeta)$ is constant, i.e.  the integrand function $F(\zeta )$ is linear with respect to $\zeta$ (this is another way of giving
the result from \eqref{pathtot1} that $\oint A_\mu dx^\mu$ vanishes up to first order in $F(\zeta)$), or if the particles travels at the speed of light (``riding the waves") along $z$ in the same direction as the waves (as $\zeta$ itself remains constant along the path for $\beta_z \rightarrow \mp 1$).  The average vector potential of the first and second halves cancels each other out also if the periodicity of the waves is such that the particle on the path encounters integer number of the wavelengths in $\Delta t$.  

\section{Aharonov-Bohm phase via the fields}
\label{AB-fields}

In this section, we work out the Aharonov-Bohm phase using the electric and magnetic fields and taking
the ``area" integrals. This is a 4D example of the usual 3D Stokes' 
theorem, where one finds $\oint {\bf A \cdot} d{\bf x} = \int {\bf B \cdot} d {\bf a}$. In the 4D case, we want 
$\oint A_\mu dx^\mu = \frac{1}{2} \int F_{\mu \nu} dx^\mu \wedge dx^\nu$, where we have written the 4D area 
$d \sigma ^{\mu \nu} \rightarrow dx^\mu \wedge dx^\nu$, using the antisymmetric wedge product. Concise details of this 
notation can be found in \cite{ryder, flanders}. Taking only the components of electric and magnetic fields given in \eqref{EB-wave}, we have 
\begin{equation}
\label{area}
\int F = 
\frac{1}{2} \int F_{\mu \nu} dx^\mu \wedge dx^\nu =  - \int \int E_x ~ dx  \wedge c dt - \int \int B_y dz \wedge dx ~.
\end{equation}
In the right hand side, we have written the tensor components in three vector notation with $E_x = F_{01}$ and $B_y = F_{13}$. The first term 
is a space-time area integral of the electric field, $E_x$, and the second term is a purely spatial area integral
of the magnetic field, $B_y$.  As in section III we will do the ``upright diamond" area and show that this is
equivalent to the loop integral of the ``upright diamond" from \eqref{upright1}. Then by boosting in the $z$ direction,
one can obtain the area intergal for the general surface from figure \ref{fig1}.
For the upright diamond loop one has a surface only in the $xt$-plane (see figure \ref{fig2}), and therefore 
only an electric contribution:
\begin{equation}
\label{diamondarea}
\int_{\rm upright} F =  - \int \int E_x ~ dx \wedge cdt   ~.
\end{equation}

We will split the upright diamond into left triangle and right triangles, evaluate these separately and then sum them up:
\begin{equation}
\int_{\rm upright} F =  - \left( \int_{\rm left} + \int_{\rm right} \right) E_x ~ dx \wedge cdt ~ . 
\end{equation}

For the upright diamond, $z=0$ and $k, \omega \rightarrow k_0 , \omega_0$, so from \eqref{EB-wave} we have  
$E_x = F_{01} = A_0 \frac{\omega}{c} f'(\omega t \pm kz) \rightarrow A_0 \frac{\omega_0}{c} f'(\omega_0 t)$. 
With this the left half of the surface integral is:
\begin{eqnarray}
\label{area-el}
- \int \int _{left} E_x ~ dx \wedge cdt &=& - \int _{-\Delta x} ^0 \int _{-x/v_0} ^{2 \Delta t + x/v_0}
A_0 \omega_0 f' (\omega_0 t) dx \wedge dt \nonumber \\
&=&  - \int _{-\Delta x} ^0  A_0  \left[f \left(2 \omega_0 \Delta t + \frac{\omega_0 x}{v_0} \right) - 
f \left(- \frac{\omega_0 x}{v_0} \right) \right] dx \\
&=& \frac{A_0 \beta_0}{k_0} \left( 2F\left(\Delta \zeta_0 \right) -  F(0) - F\left( 2\Delta \zeta_0 \right)  \right) \nonumber ~,
\end{eqnarray}
where $\Delta \zeta_0 = \omega_0 \Delta t = (k_0 / \beta_0) \Delta x$.

Likewise, for the right half of the upright diamond, we have
\begin{eqnarray}
\label{area-er}
- \int \int _{right} E_x ~ dx ~ cdt&=& - \int ^{\Delta x} _0 \int _{x/v_0} ^{2 \Delta t - x/v_0} 
A_0 \omega_0 f' (\omega_0 t) dx \wedge dt
\nonumber \\
&=&  - \int ^{\Delta x} _0  A_0 \left[f \left(2 \omega_0 \Delta t - \frac{\omega_0 x}{v_0} \right) - 
f \left( \frac{\omega_0 x}{v_0} \right) \right] dx  \\
&=& \frac{A_0 \beta_0}{k_0} \left( 2F\left(\Delta \zeta_0 \right) -  F(0) - F\left( 2\Delta \zeta_0 \right)  \right) \nonumber ~.
\end{eqnarray}

Adding \eqref{area-el} and \eqref{area-er} gives the final result for the upright diamond of
\begin{equation}
\label{uprightarea}
\int_{\rm upright} F = 2 \frac{A_0 \beta_0}{k_0} \left[ 2F(\Delta \zeta_0) -  F(0) - F(2\Delta \zeta_0)  \right] ~. \\
\end{equation}
This is the identical result for the upright diamond loop integral of equation \eqref{upright1}, which confirms the 
4D Stokes' theorem. To get the case with a general velocity in the $z$ direction, we can boost the result from
equation \eqref{uprightarea}, as we did in section III, to obtain the final result of equation \eqref{pathtot}. 

The upright diamond area integral is solely due to electric field.  However, after the boost, it gets a magnetic contribution.  
The ratio of the electric and magnetic contributions depends on the magnitude of the boost $\beta_z$.  One can calculate the electric 
and magnetic contributions separately.  Denoting the Aharonov-Bohm phase due to electric and magnetic field as $\Delta \phi_E$ 
and $\Delta \phi_B$, 
respectively, one can derive (see Appendix)
\begin{eqnarray}
\label{ABratio}
\frac{\Delta \phi_B}{\Delta \phi_E} = \frac{ \int F_{31} ~dz \wedge dx }{\int F_{10} ~dx \wedge cdt } = \pm \beta_z ~.
\end{eqnarray}

The ratio only depends on $\beta_z$.  At any speed less then $c$, the electric contribution is larger.
This is the reverse of what was found in reference \cite{brown}, under different conditions. In \cite{brown}, the time dependent
system was an infinite solenoid with a magnetic flux which was pulsed on over a short time. In this case, the magnetic Aharonov-Bohm 
phase was dominant over the phase shift due to the electric Aharonov-Bohm phase. In \cite{singleton} and \cite{singleton2}, the
case of an infinite solenoid with a slowly varying flux was considered. To linear order it was found that 
the electric and magnetic Aharonov-Bohm phase shifts were of equal magnitude but opposite sign and thus canceled. 

For the case of the electromagnetic wave, the magnetic contribution becomes equal to the electric contribution 
only when the particles travel at the speed of light. When they travel at $c$  {\it against} the waves, 
the two contributions add up; when they travel at $c$ along the waves - ``riding on the waves" - the two contributions 
cancel completely and result in a zero time-dependent Aharonov-Bohm phase, as derived in section III.  
In both cases, due to relativistic constraint, $\beta_x \rightarrow 0$ as $\beta_z \rightarrow 1$. 
This behavior is different from the case of the solenoid, where reversing the magnetic field of 
the solenoid also reverses the direction of the electric field.
Thus, in the case of the solenoid, one has the same result whether the magnetic field points along the $+z$
or $-z$ direction. In contrast, reversing the direction of the electromagnetic wave only reverses the direction of 
one of the fields, so, in one case, the electric and magnetic contributions add up, while in the other case, the 
electric and magnetic contributions tend to cancel.  

Using either the area integral of the fields or the line integral of the potentials, the Aharonov-Bohm phase picked up in 
going around the space-time loop in figure \eqref{fig1} is 
\begin{equation}
\label{ab-phase-tot}
\left(\frac{e}{\hbar c} \right) \left( \frac{2 A_0}{k'} \right) [2F(k' \Delta x ) - F (2 k' \Delta x ) - F(0) ] ~.
\end{equation}

We now recall the expansion of the term in square bracket from \eqref{pathtot}, in terms of $k' \Delta x =\Delta \zeta$, 
and apply this to \eqref{ab-phase-tot} to give the Aharonov-Bohm phase up to ${\cal O} (k' \Delta x)^3$ as
\begin{equation}
\label{ab-phase-approx}
\frac{1}{2} \left(\frac{e}{\hbar c} \right) \int F_{\mu \nu} dx^\mu \wedge dx^\nu  = \left(\frac{e}{\hbar c} \right) \oint A_\mu dx^\mu  =
-2 \left(\frac{e}{\hbar c} \right) A_0 k ' \Delta x ^2 F '' (0)   ~.
\end{equation}
From the above equation, we can see that in order for the time-dependent Aharonov-Bohm effect to be observable, one wants to have
\begin{equation}
\label{condition}
\left(\frac{e}{\hbar c} \right) A_0 k' \Delta x ^2 \sim {\cal O} (1) ~.
\end{equation}
In writing \eqref{condition} we have ignored the factor of $-2$ and assumed that $F '' (0) \sim {\cal O} (1)$, 
which would be the case if $f(\zeta_\pm)$ and $F(\zeta_\pm)$ were sinusoidal. 

We now investigate a few different scenarios for the condition in \eqref{condition}. 
In regard to the spatial size of
our loop, we could follow \cite{ford} and estimate that $\Delta x \sim 100 \mu m$, and take the speed of our particless to be 
non-relativistic with $v \sim 10 ^6 \frac{m}{s}$. Thus, from \eqref{kprime}, we have $k' \approx k \frac{c}{v} = \frac{\omega}{v}$. 
Under these assumptions, the condition in \eqref{condition} has two free parameters: $A_0$ and $\omega$. Combining all this, the condition in \eqref{condition} leads to 
\begin{equation}
\label{condition2}
A_0 \omega \sim 10^8 \frac{\mathrm{J}}{\mathrm{C} \cdot \mathrm{sec}} ~.
\end{equation}
For example, one could consider soft $x$-ray frequencies of $\omega \sim 10 ^{17}$ Hz, which would only require an amplitude of $A_0 \sim 10^{-9} \frac{J}{C}$.  As another scenario, one could consider a frequency of 
$\omega \sim 10^{14}$ Hz, which is in the infrared/optical part of the spectrum. This would correspond to the wavelength one order of magnitude less than the size of the loop considered here.  For this choice 
of $\omega$, equation \eqref{condition2} implies that one would need $A_0 \sim 10^{-6} \frac{J}{C}$. 
In general,  this sort of condition \eqref{condition2} sets a certain constraint on the experimental verification of the AB phase shift: if $A_0 \omega$ is too small, the AB-phase may be too small to detect, while if it's too large the electromagnetic force that deflects the electrons from the prescribed space-time paths may be not negligible, as the energy density is $\rho _{E\&M} \propto A_0 ^2 \omega^2$.  Some of the same conclusion were reached in \cite{ageev}.

\section{Conclusions and Summary}

In this article, we have investigated the time-dependent Aharonov-Bohm phase of a charged particle traveling around a closed space-time loop,
given in figures \ref{fig1} - \ref{fig3}, in the presence of EM plane waves traveling in the
$\pm z$ direction and with a polarization in the $x$ direction. Our work 
is a generalization of earlier works, \cite{chiao} and \cite{ford}, in that
we consider both the electric {\it and} magnetic Aharonov-Bohm effects and thus we are able to see the interplay between the two.
In addition, we calculated the time-varying Aharonov-Bohm phase both in terms of the path integral of the vector potential, 
$\oint A_\mu dx^\mu$, and in terms of the area integral of the fields, $\frac{1}{2} \int F_{\mu \nu } dx^\mu \wedge dx^\nu$. 

Our overall conclusion is that the AB phase due to EM plane waves vanishes under a broad range of conditions. Thus one must carefully choose the parameters of the electromagnetic wave and
of the loop that the electrons traverse in order to see the effect. This is similar to the conclusion of the work in 
\cite{chentsov} \cite{ageev}
where the time varying Aharonov-Bohm phase was searched for experimentally but not observed. First, from \eqref{ave-phase}, 
one finds that generally the time varying Aharonov-Bohm phase vanishes if the average of the wave form over the first half of the 
loop equals the average over the second half of the loop 
{\it i.e.} $ \langle f(\zeta) \rangle_{\rm I}  =  \langle f(\zeta) \rangle_{\rm II}$. Second, from \eqref{ab-phase-approx}, 
we showed that expanding the time varying Aharonov-Bohm phase to the lowest order gave a phase with a magnitude of 
$\left(\frac{e}{\hbar c} \right) A_0 k ' \Delta x ^2 $. In this long wavelength approximation, 
only the second order term in $k' \Delta x$ remained - the zeroth and first order terms vanished.  This 
smallness of the time varying Aharonov-Bohm phase is similar to earlier results in a different context \cite{singleton, singleton2, singleton3}.  Third, AB phase vanishes when the velocity of the particle traversing the loop moves completely in the $z$ direction (``riding along") at the speed of light, $\beta_z =1$.

In order to see the time varying Aharonov-Bohm effect, we found a general condition under the perturbative approximation, equation \eqref{condition}. This condition depended on the wave amplitude, frequency, particle velocity and
size of the loop - $A_0$, $\omega$, $v$ and $\Delta x$. By making some reasonable choices for the speed of the particles and the loop size ({\it i.e.} $v \sim 10^6 \mathrm{m}/\mathrm{s}$ and $\Delta x \sim 100 \mu$m), we arrived at a more specific condition for the frequency and amplitude of the EM waves: 
$A_0 \omega \sim 10^8 \mathrm{J} / (\mathrm{C} \cdot \mathrm{sec})$.  Due to the electromagnetic forces on the particle, this constraint may set a certain limitation on the experimental verification of the AB effect under the perturbative approximation.\\

{\par\noindent {\bf APPENDIX I}}
\\
In this appendix, we derive the ratio $\Delta \phi_B / \Delta \phi_E$ given in \eqref{ABratio}.  The total 
phase shift for the tilted diamond, given in \eqref{pathtot}, can be split into an electric and magnetic contribution as 
\begin{eqnarray}
\label{dp}
\Delta \phi = \Delta \phi_E + \Delta \phi_B =  \frac{2A_0}{k'}  \left[ 2F(\Delta \zeta) -  F(0) - F(2\Delta \zeta) \right] ~.
\end{eqnarray}
To find the ratio of the magnetic to electric contributions, we just need to find $\Delta \phi_E$ for the ``tilted diamond" and then use 
this in \eqref{dp} to obtain $\Delta \phi_B$. As in section IV, we calculate the electric part of the tilted diamond by
calculating the left and right sides and adding these together. Using \eqref{EB-wave}, the electric contribution for the left side of the
tilted diamond is 
\begin{eqnarray}
\label{dpl}
(\Delta \phi_E )_{left} &=& - \int \int_{left} E_x ~ dx \wedge cdt \nonumber \\
&=& -A_0 \frac{\omega}{c} \int ^0 _ {- \Delta x} dx \left( \int_{-x/v_x}^{2 \Delta t + x/v_x} f'(\omega t \pm kz) c dt \right)    \\
&=& -A_0 \frac{\omega}{\omega'} \int ^0 _ {- \Delta x} \left[ f\left( 2 \omega ' \Delta t +\omega ' x / v_x \right) 
-f \left( -\omega ' x / v_x \right) \right] ~dx  \nonumber ~.
\end{eqnarray}
In the above, we have replaced $z= v_z t$ using \eqref{path1} and defined 
$\omega ' =(\omega \pm k v_z) = \omega (1 \pm \beta_z)$. Next, performing the $x$ integration in \eqref{dpl}, we find 
\begin{eqnarray}
\label{dpl2}
(\Delta \phi_E )_{left} &=& - A_0 \frac{\omega v_x}{(\omega')^2 } \left[F(2 \omega' \Delta t) -F(\omega' \Delta t) 
+F(0) - F(\omega' \Delta t) \right] \nonumber \\
&=& A_0 \frac{\beta_x}{k(1\pm \beta_z)^2} \left[ 2 F(\omega' \Delta t) - F(2 \omega' \Delta t) - F(0) \right]   \\
&=& A_0 \frac{1}{k' (1 \pm \beta_z)} \left[ 2 F(\Delta \zeta) - F(2 \Delta \zeta) - F(0) \right]  \nonumber ~,
\end{eqnarray}
where we have used $k = \frac{\beta_x}{1 \pm \beta_z}k'$ from \eqref{kprime} and $\Delta \zeta = \omega' \Delta t$
from \eqref{zeta}. In the same way, one can obtain the right half of the tilted diamond,  
\begin{eqnarray}
\label{dpr}
(\Delta \phi_E )_{right} &=& - \int \int_{right} E_x ~ dx \wedge cdt \nonumber \\
&=& -A_0 \frac{\omega}{c} \int _0 ^ { \Delta x} \left( \int_{x/v_x}^{2 \Delta t - x/v_x} f'(\omega t \pm kz) c dt \right) dx   \\
&=& -A_0 \frac{\omega}{\omega'} \int _0 ^{\Delta x} \left[ f\left( 2 \omega ' \Delta t - \omega ' x / v_x \right) 
-f \left( \omega ' x / v_x \right) \right] ~dx  \nonumber ~.
\end{eqnarray}
Next, performing the $x$ integration in \eqref{dpr}, we find 
\begin{eqnarray}
\label{dpr2}
(\Delta \phi_E )_{right} &=& - A_0 \frac{\omega v_x}{(\omega')^2 } \left[F(2 \omega' \Delta t) -F(\omega' \Delta t) 
+F(0) - F(\omega' \Delta t) \right] \nonumber \\
&=& A_0 \frac{\beta_x}{k(1\pm \beta_z)^2} \left[ 2 F(\omega' \Delta t) - F(2 \omega' \Delta t) - F(0) \right]   \\
&=& A_0 \frac{1}{k' (1 \pm \beta_z)} \left[ 2 F(\Delta \zeta) - F(2 \Delta \zeta) - F(0) \right]  \nonumber  ~.
\end{eqnarray}
Adding the left and right sides together, we get the total electric contribution for the tilted diamond as 
\begin{equation}
\Delta \phi_E = \frac{2A_0}{k' (1\pm \beta_z)} \left[ 2F(\Delta \zeta) - F(0) - F(2\Delta \zeta) \right] ~.
\end{equation}
From \eqref{dp}, this gives for the magnetic contribution
\begin{equation}
\Delta \phi_B = \Delta \phi - \Delta \phi _E =
\frac{\pm 2A_0 \beta_z}{k' (1\pm \beta_z)} \left[ 2F(\Delta \zeta) - F(0) - F(2\Delta \zeta) \right] ~.
\end{equation}
Hence, 
\begin{equation}
\frac{\Delta \phi_B}{\Delta \phi_E} =  \pm \beta_z ~,
\end{equation}
which is the result given in \eqref{ABratio}.
\\

{\par\noindent {\bf Acknowledgments:}} DS is supported by a 2015-2016 Fulbright Scholars Grant to Brazil. 
DS wishes to thank the ICTP-SAIFR in S{\~a}o Paulo for it hospitality. DS also acknowledges support by a grant 
(number 1626/GF3) in Fundamental Research in Natural Sciences by the Science Committee of the Ministry of Education 
and Science of Kazakhstan.  AY wishes to thank the members of Deptartment of Physics, California State University, Fresno, for thier hospitarity during AY's stay while this work was underway.\\


\begin{thebibliography}{99}

\bibitem{AB} Y Aharonov and D. Bohm, Phys. Rev. {\bf 115}, 484 (1959).

\bibitem{ES} W. Ehrenberg and R. E. Siday, Proc. Phys. Society B {\bf 62}, 8 (1949).

\bibitem{chambers} R. G. Chambers, Phys. Rev. Lett. {\bf 5}, 3 (1960).

\bibitem{tonomura} A. Tonomura, et al., Phys. Rev. Lett. {\bf 56}, 792 (1986).

\bibitem{kampen} N.G. van Kampen, Phys. Lett. {\bf A106}, 5 (1984).

\bibitem{roy} S. M. Roy and V. Singh, Nuovo Cimento A {\bf 79}, 391 (1984). 

\bibitem{brown} R. A. Brown and D. Home, Nuovo Cimento B {\bf 107}, 303 (1991).

\bibitem{gaveau} B. Gaveau, A.M. Nounou, and L.S. Schulman, Found. Phys. {\bf 41}, 1462 (2011). 

\bibitem{singleton} D. Singleton and E. Vagenas, Phys. Lett. B {\bf 723}, 241 (2013).

\bibitem{singleton2} J. MacDougall and D. Singleton, J. Math. Phys. {\bf 55}, 042101 (2014).

\bibitem{chiao} B. Lee, E. Yin, T. K. Gustafson, and R. Chiao,  Phys. Rev. {\bf A45}, 4319 (1992). 

\bibitem{ford} J. Hsiang and L. H. Ford, Phys. Rev. Lett. {\bf 92}, 250402 (2004).

\bibitem{marton} L. Marton, J. A. Simpson, and J. A. Suddeth, Rev. Sci. Instr. {\bf 25}, 1099 (1954).

\bibitem{brill} F.G.Werner, D.Brill, Phys. Rev. Lett. {\bf 4}, 344 (1960).

\bibitem{singleton3} J. Macdougall, D. Singleton, and E. C. Vagenas, Phys. Lett. {\bf A379}, 1689 (2015).

\bibitem{chentsov} Yu. V. Chentsov, Yu. M. Voronin, I. P. Demenchonok,
and A. N. Ageev, Opt. Zh. {\bf 8}, 55 (1996).

\bibitem{ageev} A. N. Ageev, S. Yu. Davydov, and A. G. Chirkov, 
Technical Phys. Letts. {\bf 26}, 392 (2000).

\bibitem{AC} Y. Aharonov and A. Casher, Phys. Rev. Lett. {\bf 53}, 319 (1984).

\bibitem{bright} M. Bright and D. Singleton, Phys. Rev. {\bf D91}, 085010 (2015).

\bibitem{coleman} S. Coleman, Phys. Lett. {\bf 70B}, 59 (1977).

\bibitem{ryder} L. Ryder, {\it Quantum Field Theory}, 2$^{nd}$ ed. (Cambridge University Press, Cambridge, UK, 1996), Secs. 2.9.

\bibitem{flanders} H. Flanders, {\it Differential Forms with Applications to the Physical Sciences}, (Dover Publications Inc., New York, USA, 1989).

\end{thebibliography}
\end{document}